\documentclass[10pt]{article}
\usepackage{amsmath}
\usepackage{amssymb}
\usepackage{graphicx}
\usepackage{cite}
\usepackage{color}
\usepackage[labelfont=bf,labelsep=period,justification=raggedright]{caption}
\makeatletter
\renewcommand{\@biblabel}[1]{\quad#1.}
\makeatother

\date{}

\begin{document}

\begin{flushleft}
{\Large
\textbf{Resolution of the stochastic strategy spatial prisoner's dilemma by means of particle swarm optimization}
}\sffamily
\\[3mm]
\textbf{Jianlei Zhang,$^{1,\ast}$ Chunyan Zhang,$^{1}$ Tianguang Chu,$^{1,2,\dagger}$ Matja{\v z} Perc$^{3,\S}$}
\\[2mm]
{\bf 1} State Key Laboratory for Turbulence and Complex Systems, College of Engineering, Peking University, Beijing, China
{\bf 2} Key Laboratory of Machine Perception, Ministry of Education, Peking University, Beijing, China
{\bf 3} Faculty of Natural Sciences and Mathematics, University of Maribor, Slovenia
\\[1mm]
$^{\ast}$jianleizhang@pku.edu.cn\\
$^{\dagger}$chutg@pku.edu.cn\\
$^{\S}$matjaz.perc@uni-mb.si [www.matjazperc.com]
\end{flushleft}
\sffamily
\section*{Abstract}
We study the evolution of cooperation among selfish individuals in the stochastic strategy spatial prisoner's dilemma game. We equip players with the particle swarm optimization technique, and find that it may lead to highly cooperative states even if the temptations to defect are strong. The concept of particle swarm optimization was originally introduced within a simple model of social dynamics that can describe the formation of a swarm, i.e., analogous to a swarm of bees searching for a food source. Essentially, particle swarm optimization foresees changes in the velocity profile of each player, such that the best locations are targeted and eventually occupied. In our case, each player keeps track of the highest payoff attained within a local topological neighborhood and its individual highest payoff. Thus, players make use of their own memory that keeps score of the most profitable strategy in previous actions, as well as use of the knowledge gained by the swarm as a whole, to find the best available strategy for themselves and the society. Following extensive simulations of this setup, we find a significant increase in the level of cooperation for a wide range of parameters, and also a full resolution of the prisoner's dilemma. We also demonstrate extreme efficiency of the optimization algorithm when dealing with environments that strongly favor the proliferation of defection, which in turn suggests that swarming could be an important phenomenon by means of which cooperation can be sustained even under highly unfavorable conditions. We thus present an alternative way of understanding the evolution of cooperative behavior and its ubiquitous presence in nature, and we hope that this study will be inspirational for future efforts aimed in this direction.

\section*{Introduction}
Cooperation is the basis for complex organizational structures in biological as well as social systems. Nevertheless, understanding the emergence and stability of cooperative behavior in the context of Darwinian selection remains a challenge to date. The dilemmas of cooperation are usually tackled within the framework of evolutionary game theory \cite{hofbauer_98, nowak_06, sigmund_10}. Although several mechanism allowing for the evolution of cooperation have already been identified \cite{nowak_s06}, the resolution of social dilemmas and the closely related avoidance of the ``tragedy of the commons'' \cite{hardin_g_s68} is still considered an open problem. The prisoner's dilemma game \cite{axelrod_84}, in particular, has attracted considerable attention in the past three decades \cite{doebeli_el05, szabo_pr07, roca_plr09, perc_bs10}, and to date it is widely consider as a paradigmatic example for the tensions between social welfare and individual interests \cite{santos_prl05, szabo_pre05, perc_njp06a, tomassini_ijmpc07, gomez-gardenes_prl07, szolnoki_epl07, chen_xj_pre08b, poncela_ploso08, pestelacci_bt08, fu_pre09, poncela_epl09, van-segbroeck_prl09, szabo_epl09, szabo_pre09, pena_pre09, wu_zx_pre09, poncela_njp09, wu_b_pone10, rong_pre10, cardillo_njp10, szolnoki_pre10b, fu_jtb10, perc_pone10}. Cooperation and defection are the two strategies that are at the heart of the prisoner's dilemma game. In general, while cooperators sacrifice some of their personal fitness for the benefit of the society, defectors succumb to the temptations and take full advantage of them. The prisoner's dilemma captures this situation by means of the following payoffs: mutual cooperation yields the reward $R$, mutual defection leads to punishment $P$, and the mixed choice gives the cooperator the sucker's payoff $S$ and the defector the temptation $T$. The payoff ranking thus satisfies $T>R>P>S$. In the iterated prisoner's dilemma game the assumption that the mutual cooperation yields the highest collective income imposes another constraint, namely $2R>T+S$. This makes it clear that the rational (selfish) action is to defect, and according to the fundamental principles of Darwinian selection, cooperation extinction is inevitable. Full defection is indeed the only stable Nash equilibrium for the prisoner's dilemma game in well-mixed populations.

Since the seminal paper by Nowak and May \cite{nowak_n92b}, however, we know that this may not be the case for spatial interactions. Although not universally applicable \cite{hauert_n04}, spatial reciprocity is recognized as a potent promoter of cooperative behavior, even more so on complex networks \cite{abramson_pre01, zimmermann_pre04, santos_pnas06, poncela_njp07, perc_njp09} (for a comprehensive review see \cite{szabo_pr07}). Other prominent mechanism promoting cooperation are kin selection \cite{hamilton_jtb64}, direct and indirect reciprocity \cite{nowak_jtb98, brandt_pnas05, brandt_jtb06}, as well as group selection \cite{dugatkin_bs96, traulsen_pnas06, szolnoki_njp09}, to name but a few.

Inspired by previous works on this subject, we here introduce particle swarm  optimization \cite{kennedy_ieee95, eberhart_is95, eberhart_96} to the players engaging in the prisoner's dilemma game on a square lattice \cite{szabo_pre98}, with the aim of investigating its impact on the evolution of cooperation. However, we abandon the commonly considered assumption that the players can choose only between the two pure strategies, namely to either cooperate or to defect. Real-life situations are often more complex than that, and indeed there is a lot of gray between the black and white extremes. Motivated by this fact, we here consider stochastic strategies, such that the cooperativeness of each players is determined by $W \in [0,1]$. $W=1$ returns full cooperation, while $W=0$ returns full defection. These are the two extremes recovered from our present setup. Between $0<W<1$, however, there exists a continuous set of strategies that can be considered either as predominantly cooperative (if $W>0.5$) or predominantly defective (if $W<0.5$). Moreover, while the evolution of strategies is traditionally performed by means of different strategy adoption (or updating) rules (see \cite{szabo_pr07} for a comprehensive review), we here take a much less explored avenue, namely by considering the aforementioned particle swarm optimization as the driving force behind strategy evolution. The particle swarm optimization algorithm is based on a simplified social model that is tightly tied to the theory of swarming \cite{kennedy_ieee95, eberhart_is95, eberhart_96}. A traditional analogy is a swarm of bees searching for a food source. In this analogy, each bee (considered here as a particle) makes use of its own memory as well as the knowledge obtained by the swarm as a whole, to find the best available food source. Particle swarm optimization can also be considered as being representative for multidimensional search (for example to find an optimum of a utility function). Typically, a number of simple entities (the ``particles'') is randomly positioned in the search space, and to each a velocity vector is assigned, which is subsequently used to update the current position of each particle in the swarm. Each particle then proceeds by evaluating the objective function at its current location, and finally to determining its movement through the search space by combining some aspects of the history of its own current as well as other potentially optimal locations with those of one or more members of the swarm. Thus, the process makes use of the memory of each particle, as well as the knowledge gained by the swarm as a whole. The next iteration takes place after all the particles have moved once. Eventually the swarm, like a flock of birds collectively foraging for food, is likely to move closer to an optimum of the utility function. Accordingly, the particles (bees, birds, players) therefore should have a tendency to fly towards better and better areas over the course of the search process.

Here we focus specifically on introducing the particle swarm optimization algorithm to the strategy updating process in the stochastic strategy prisoner's dilemma game on the square lattice. In agreement with the above described general concept, each individual is assigned a variable from the unit interval determining its level of cooperativeness (or willingness to cooperate). Likewise, a velocity vector is assigned to every player. Following this initialization, each player makes use of its own memory (i.e., keeping score of the most profitable individual strategy in the past), as well as use of the knowledge gained by the swarm (i.e., the nearest neighbors) as a whole, to find the best available strategy for itself and the society. In particular, the particle swarm optimization algorithm makes use of the velocity vector to update the current strategy of each player in the swarm. In this sense our study can be considered related to previous works investigating the effects of mobility on the evolution of cooperation \cite{vainstein_pre01, vainstein_jtb07, meloni_pre09, helbing_pnas09, droz_epjb09, jiang_ll_pre10}, although it relies on an essentially different algorithm. The outline of the latter is as follows: 1) Start with a set of strategies (i.e., cooperation probabilities $W$) that are initially uniformly distributed in the $[0,1]$ interval. 2) Calculate a velocity vector for each strategy in the swarm. 3) Update the strategy of each agent, using its previous value and the updated velocity vector. 4) Go to step $2$ and repeat until convergence. All the details of this setup are described in the Methods section, while here we proceed with presenting the main results.

\section*{Results}
\begin{figure}
\begin{center}\includegraphics[width=9cm]{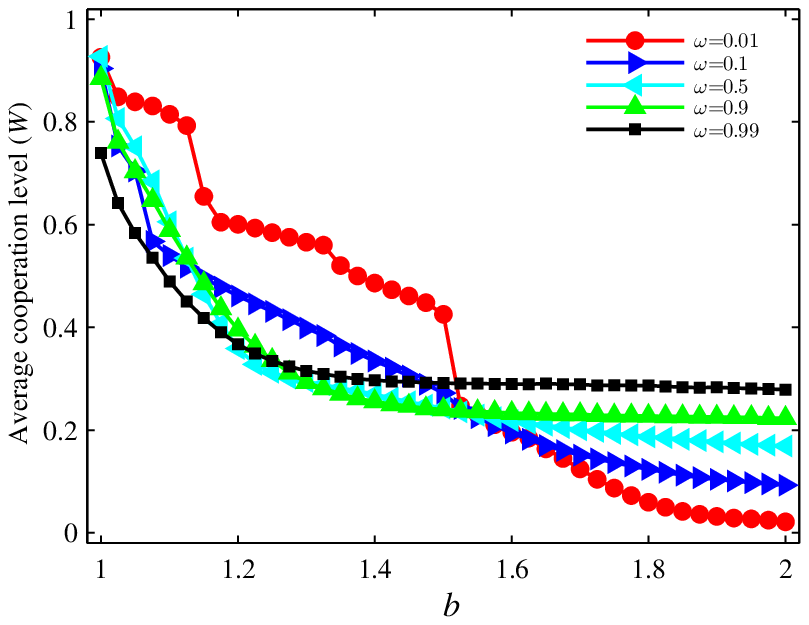}\end{center}
\caption{\textbf{Average level of cooperation in dependence on $b$ for different values of $\omega$.} It can be observed that while imitating the best performing player in the swarm ($\omega \to 0$) might be beneficial at low temptations to defect, imitating personal success ($\omega \to 1$) is definitively better for the evolution of cooperation in strongly defection-prone environments. Each data point is an average of the final outcome (stationary state) of the game over $100$ independent realizations. Lines connecting the symbols are just to guide the eye.}
\label{fig1}
\end{figure}

We start by presenting the average level of cooperation, defined as $N^{-1}\sum_i W(i)$ where $N$ is the system size and $i$ runs over all the players in the population, in dependence on the temptation to defect $b$ for different values of $\omega$ (for the definition see the Methods section) in Fig.~\ref{fig1}. Expectedly, the average level of cooperation decreases as $b$ increases for all $\omega$. However, while for $\omega \to 0$ the cooperative behavior dies out completely at high values of $b$, for $\omega \to 1$ the average level of cooperation hovers comfortably over $1/3$, even when the maximal $b=2$ limit is reached. For intermediate and low values of $b$, however, small values of $\omega$ may yield overall higher average levels of cooperation. It is thus intriguing to find that the introduced particle swarm optimization in the strategy updating, fine-tuned by means of the parameter $\omega$, can be responsible for the emergence of cooperative behavior across the whole span of defection temptation values, as well as for its dominance at low values of $b$. More precisely, two regimes can be differentiated. For $b<1.5$ intermediate and high values of $\omega$ are actually detrimental for the evolution of cooperation, while for $b>1.5$ the higher the $\omega$ the higher the stationary level of cooperative behavior. These results make it clear that low $\omega$ (e.g., $\omega=0.01$) strongly support the cooperation level for small $b$, up to $b \simeq 1.2$, whereas high $\omega$ are much better suited for cooperation to evolve under this dynamics in strongly defection-prone environments. At this point we argue that for $\omega \to 1$, when players imitate their best past actions rather than the best players in the swarm (see Methods for details), the proposed strategy updating rule warrants the most significant benefits to cooperative behavior if looking at the entire range of $b$ values, thus in turn resolving the prisoner's dilemma.

\begin{figure}
\begin{center}\includegraphics[width=12cm]{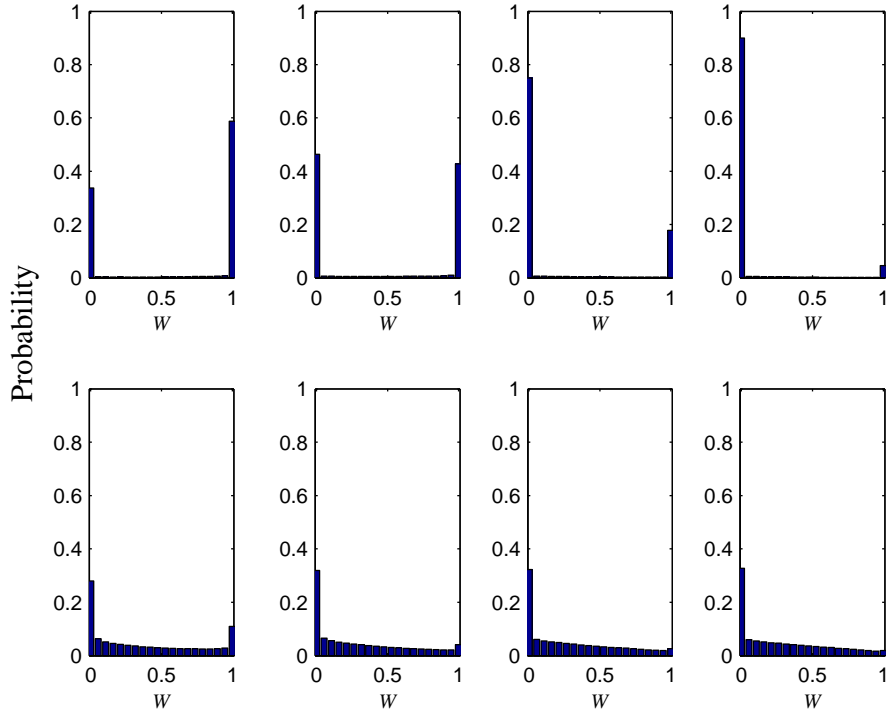}\end{center}
\caption{\textbf{Distribution of strategies in the whole population, as obtained for different combinations of $b$ and $\omega$.} It can be observed that for $\omega=0.01$ the nature of the stochastic strategy prisoner's dilemma game is essentially completely overridden by the selfish drive of players to reach the highest current payoffs in the swarm, in turn virtually completely transforming the game to its two-strategy [only $W=0$ (full defection) or $W=1$ (full cooperation) strategies are present in the population] version. Conversely, for $\omega=0.99$ the full spectrum of available strategies is exploited to arrive at the final stationary state. Note that the horizontal axis displays the willingness to cooperate $W$ (defining the strategy of every player), while the vertical axis depicts the probability that this strategy is present in the population. Depicted results are averages of the final outcome (stationary state) over $100$ independent realizations.}
\label{fig2}
\end{figure}

In order to obtain an understanding of these results, we first systematically analyze the impact of $\omega$ on the final distribution of strategies in the whole population for various values of $b$, as depicted in Fig.~\ref{fig2}. Note that for $\omega=0.01$ the distribution of strategies is very monotonous, while for $\omega=0.99$ much more diversity is inferable. Both observations are virtually independent of $b$. Since the parameter $\omega \in [0,1]$ determines the tendency of every player to either adopt the most profitable strategy in its past actions ($\omega \to 1$) or the strategy of the most successful player in its neighborhood ($\omega \to 0$), these results can be understood very well. In particular, for $\omega=0.01$ individuals are strongly inclined to imitate the best-performing strategies in the swarm, irrespective of their personal experience in the past. This narrow-sightedness inevitably results in strongly polarized distributions, as only either pure cooperators or pure defectors are the ones most likely to have the overall highest payoffs. Note that this is because the payoffs are directly scaled by $W$ (see Methods). Conversely, for $\omega=0.99$ the situation is very different since players will focus on their own past actions and learn from them in order to arrive at the best possible strategy. This has the advantage that, unlike for $\omega=0.01$, here only the immediate neighborhood is explicitly taken into account. For high values of $b$ local considerations are obviously much more important than for low values of $b$. In the latter case, the nearest neighbors can much easily be neglected since the environment on its own is not strongly favorable for defectors, and hence cooperators can prevail even if overlooking the detailed distribution of strategies in their immediate neighborhood. An additional advantage of small $\omega$, however, is that by focusing only (or predominantly) on the best-performing players in the swarm, the average level of cooperativeness can be maximized more efficiently (as evidenced by results presented in Fig.~\ref{fig1}). But if the temptation to defect is strong the strictly local considerations are much more important, as proper adaptation is then crucial for cooperators to survive. Accordingly, for high values of $b$ higher $\omega$ yield better results (higher average level of cooperation) by exploiting effectively the whole array of available strategies to respond properly (\textit{locally} properly) to invading defectors. At low values of $b$, however, these locally optimal adaptations (warranted by $\omega \to 1$) might be less effective than the more globally inspired actions (warranted by $\omega \to 0$).

\begin{figure}
\begin{center}\includegraphics[width=11cm]{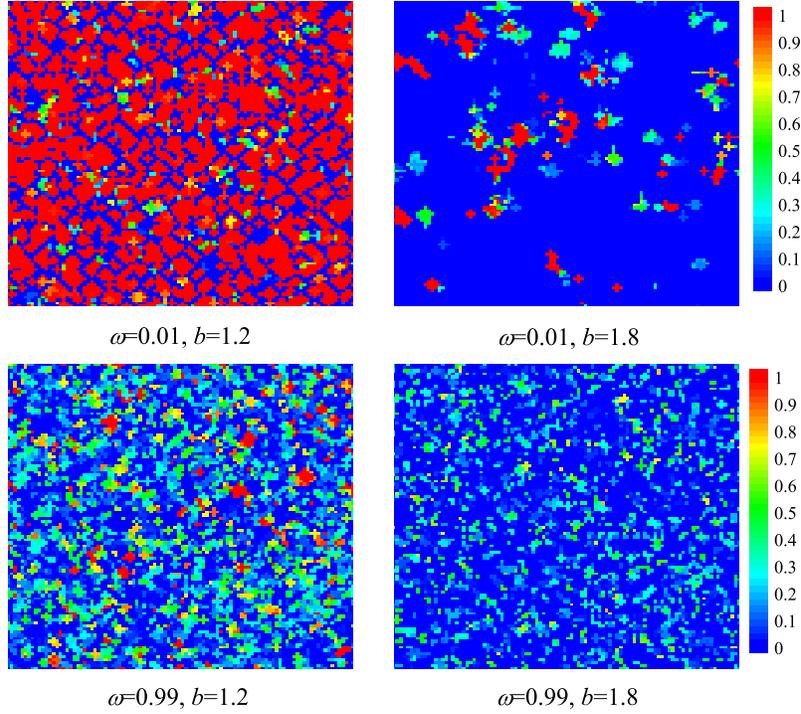}\end{center}
\caption{\textbf{Characteristic spatial distributions of strategies, as obtained for different combinations of $b$ and $\omega$.} As concluded from results depicted in Fig.~\ref{fig2}, for low values of $\omega$ only the two ``extreme'' strategies (with rare exceptions) are adopted, while for high values of $\omega$ the whole array of available strategies comes into play. Moreover, it is interesting to observe that values of $\omega \to 0$ yield the well-known clustering of cooperators \cite{nowak_n92b} on the square lattice, while the snapshots for $\omega=0.99$ seem to have these feature somewhat less pronounced, although still clearly inferable (note that the distinction of clusters is somewhat difficult due to the continuous array of possible strategies). This suggests that, besides the clustering of cooperators, additional mechanisms may underlie the survival of cooperators at high temptations to defect and $\omega \to 1$ within the present setup. The color encoding, as depicted right, indicates the values of $W$ for each individual player.}
\label{fig3}
\end{figure}

\begin{figure}
\begin{center}\includegraphics[width=11cm]{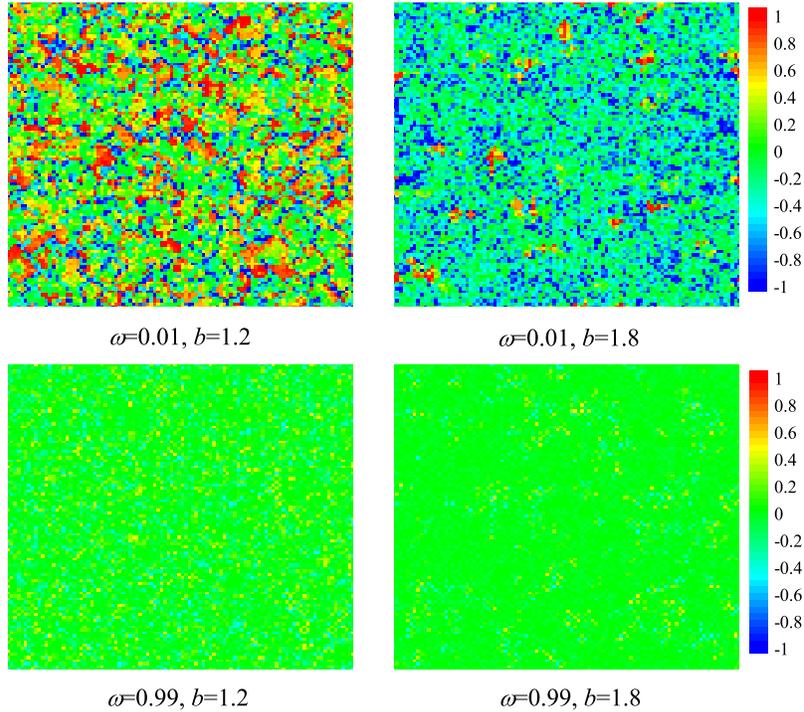}\end{center}
\caption{\textbf{Characteristic spatial distributions of velocities, as obtained for different combinations of $b$ and $\omega$.} Top row depicts results for $\omega=0.01$, while bottom row features results for $\omega=0.99$. Irrespective of $b$, it can be observed that for $\omega=0.99$ the whole population essentially becomes a swarm in that the velocities of all players are much the same and close to zero. The fact that the prevailing velocity is close to zero simply reflects that the stationary state has been reached by means of adaptive, locally-inspired and slow strategy changes (which are, however, very effective even if the temptations to defect are strong). For $\omega=0.01$, however, only isolated clusters can be considered to act as swarms, while the majority of players cannot be associated with any kind of group dynamics and is simply caught in the futile pursuit for the highest, yet for the majority unattainable, payoffs. These results indicate that swarming is an important agonist that promotes cooperation at high temptations to defect (see results presented in Fig.~\ref{fig1}). The color encoding, as depicted right, indicates the values of $V_{i,n}$ for each individual player, where $n$ was chosen sufficiently large such that the stationary state of the game has been reached. Importantly, we note that for $\omega=0.01$ the stationary state has in fact been reached, although at a given instance in time the average velocity in the population might be different from zero.}
\label{fig4}
\end{figure}

These conclusions can be corroborated further by examining characteristic snapshots of strategy and velocity distributions for key combinations of $b$ and $\omega$, as presented in Figs.~\ref{fig3} and \ref{fig4}. Focusing first on the distribution of strategies in Fig.~\ref{fig3}, it can be inferred that for $\omega=0.01$, where only the most successful strategies within the whole swarm can spread rapidly due to the workings of the particle swarm optimization algorithm, the strategy distribution becomes very monotonous, leading to the isolation of homogeneous groups of players characterized either by $W=0$ or $W=1$, respectively. This holds irrespective of $b$, only that for strong temptations to defect the clusters of strongly cooperative players become rarer. Note that in this parameter region the here studied stochastic strategy prisoner's dilemma game actually becomes strikingly similar to the classical two-strategy spatial prisoner's dilemma game \cite{nowak_n92b, szabo_pre98}, where the clustering of cooperators is the main driving force prohibiting the full dominance of defectors. Conversely, for $\omega=0.99$, where the particle swarm optimization algorithm is driven by the past experience of every individual player (rather than the swarm as a whole), highly heterogeneous kaleidoscopes appear, and it is indeed this diversity that warrants a high level of cooperativeness even by strong temptations to defect. In particular, snapshots in the bottom panel of Fig.~\ref{fig3} indicate that many clusters consist of a small amount of players with a high cooperation level (i.e., $W$ close to $1$), surrounded by players with comparatively lower $W$ values. This in turn implies that not the clustering itself is crucial for the sustenance of cooperation, but actually the aggregation of such clusters itself, which enables the players with higher cooperation level to survive the evolutionary process. Note that the high cooperation level within clusters provides surrounding individuals with a safe source of benefits that are sufficient to resist the invasion of predominantly defective (i.e., $W$ close to $0$) players. The particle swarm optimization algorithm thus spontaneously generates the diversity needed for cooperation to survive at high $b$, much by means of the same mechanism that was reported previously for manually introduced heterogeneous states \cite{perc_pre08}. Of course, players located in the interior of such clusters enjoy the benefits of mutual cooperation and are therefore able to survive despite the constant exploitation by defectors, yet this positive effect is additionally amplified by the diversity and the hierarchical local structures that give additional strength to the cooperative strategy, while at the same time provide no benefits for defectors.

Moreover, by examining the characteristic distributions of velocities presented in Fig.~\ref{fig4}, we can obtain further insight with regards to the evolution of the strategies and their adaptation. Note that by means of Eqs.~(1) and (2) (see the Methods section), the two quantities are strongly interdependent. For $\omega=0.01$, even though the snapshots are taken in the stationary state (where the average level of cooperation is stable), the majority of players will have the velocity very different from $0$ (although on average over time and space it is virtually zero, thus assuring the stationary state being reached). This indicates that players will constantly try to reach the currently maximal payoff in the swarm, despite the fact that for the majority this will be unattainable. The locally high velocity values also indicate that the evolutionary process at low values of $\omega$ is quite violent and fast, with the population therefore unable to cope with high temptations to defect. Conversely, for $\omega=0.99$ the situation is very different. Here the majority of players will adapt their strategy very slowly to the changing local influences, which yields the velocity profile for every player being very close to zero. These conclusions are valid practically irrespective of $b$ for the two considered values of $\omega$, but the average level of cooperation is in fact very much different. While individually optimal past strategies in the particle swarm optimization algorithm yield a slow but stable and very effective response even to severe defector attacks, population-wide (or swarm-wide) pursuit for extraordinary benefits proves insufficiently effective to sustain cooperative behavior at high $b$ values. The latter approach, however, may be superior at low temptations to defect, where local considerations are not so vital, and where the pursuit of individual benefits can be successful even if driven by globally-inspired fast and bold actions.

\section*{Summary}
In sum, we have studied the impact of particle swarm optimization on the evolution of cooperation in the stochastic strategy spatial prisoner's dilemma game. The strategy updating was guided by the particle swarm optimization algorithm, using as input the individual memory of every player (i.e., keeping score of the most profitable individual strategy in the past) as well as the knowledge gained by the swarm (i.e., the nearest neighbors) as a whole. By means of extensive simulations, we found that cooperative behavior can prevail in large regions of the parameter space defining the stochastic strategy prisoner's dilemma game, thus effectively leading to the resolution of the dilemma in favor of pro-social behavior. In particular, we have demonstrated that imitating the most profitable strategy in the swarm may lead to full dominance of cooperation at moderate temptations to defect, while imitating the best individual actions in the past may lead to the survival of cooperative behavior even if the environment is strongly prone to defection. We have also investigated the actual strategy configurations in the population as well as pertaining spatial distributions of strategies and velocities, for which we have found to be closely tied to the setup of the particle swarm optimization algorithm, and in fact instrumental for the understanding of the observed promotion of the evolution of cooperation. We hope that our work will offer new ways of ensuring cooperation in situations constituting a social dilemma, and that it will be an inspiration for future research when considering the very interesting combination of intelligent algorithms and evolutionary games.

\section*{Methods}
We consider an evolutionary stochastic strategy prisoner's dilemma game on a square lattice, consisting of $100 \times 100$ players with nearest-neighbor interactions and periodic boundary conditions. Initially the strategies of all players are drawn randomly from uniformly distributed values of $W$ in the $[0,1]$ interval, whereby $W$ determines the cooperativeness of each individual (or the willingness to cooperate). While $W=1$ returns full cooperation and $W=0$ returns full defection, between $0<W<1$ there exists a continuous set of strategies that can be considered either as being predominantly cooperative (if $W>0.5$) or predominantly defective (if $W<0.5$), hence constituting a stochastic strategy version of the prisoner's dilemma game.

Players interact pairwise with all their nearest neighbors, thereby receiving payoffs that can be summarized succinctly by the rescaled payoff matrix
\[\begin{array}{*{20}{c}}
   {} & {\begin{array}{*{20}{ll}}
    \ \ \ \ \ \ \ \ \ \ \ \ \ C & \ \ \ \ \ \ \ \ \ \ \ \ \ D  \\
\end{array}}  \\
   {\begin{array}{*{20}{c}}
   C \!\!\!\!\!\! \\
   D  \!\!\!\!\!\!\\
\end{array}} & {\left( {\begin{array}{*{20}{c}}
   {W(i)*W(j)} &\ {0}  \\
   b*W(j)*(1-W(i)) & \ 0  \\
\end{array}} \right)}  \\
\end{array}\]
where $W(i)$ and $W(j)$ define the level of cooperativeness of players $i$ and $j$, respectively. This setup entails $b$ as the only free parameter determining the temptation to defect, but it is well-known that the essence of the prisoner's dilemma game is thereby left intact \cite{nowak_n92b}.

The stochastic strategy prisoner's dilemma game is iterated forward in time using a synchronous Monte Carlo updating scheme. First, each player accumulates its payoff by playing the game with all four of its nearest neighbors. Subsequently, players have to decide what strategy they will adopt in the next round (i.e., what will their new $W(i)$ be), which we here determine by means of the particle swarm optimization algorithm. Its implementation is simple and intuitive, as follows. Initially, at time step $n=0$, all players are assigned the same velocity $V_{i,n}=0$. For each following $n$, the velocity vector $V_{i,n}$ of every player $i$ is updated according to
\begin{equation}
V_{i,n+1}=V_{i,n}+\omega[W(i,h)-W(i,n)]+(1-\omega)[W(\star,n)-W(i,n)],
\end{equation}
and the strategy follows directly as
\begin{equation}
W(i,n+1)=W(i,n)+V_{i,n+1},
\end{equation}
where in Eq.~(1) $W(i,h)$ is the most profitable strategy of player $i$ in all its past actions, whereas $W(\star,n)$ is the best performing strategy in the swarm (here considered to be composed of the four nearest neighbors). The parameter $\omega \in [0,1]$ determines the tendency of every player to either adopt the most profitable strategy in its past actions or the current strategy of the most successful player within the swarm. In particular, $\omega=1$ implies that the player will definitely imitate its past best action, i.e., the strategy that in the past yielded the highest payoff. On the other hand, $\omega=0$ implies that the player will copy the strategy of the currently best performing player in its neighborhood. Intermediate values of $\omega$ interpolate linearly between these two extremes. Besides the temptation to defect $b$, $\omega$ is here considered as the second crucial system parameter.

\clearpage

\end{document}